\documentclass[aps,preprint,showpacs]{revtex4}

\usepackage{graphicx,epsfig}
\usepackage{amssymb}
\begin{document}

\title{ Scattering and absorption sections of  nonlinear electromagnetic black holes}

\author{J.C.  Olvera$^1$ }
\email{jcolvera@fis.cinvestav.mx}
\author{L. A. L\'opez $^2$}
\email{lalopez@uaeh.edu.mx}

 \affiliation{$^1$ Dpto de F\'isica, Centro de Investigaci\'on y de Estudios Avanzados del I.P.N,
14-740 Apdo, DF, M\'exico}
\affiliation{$^2$  \'Area Acad\'emica de Matem\'aticas y F\'isica.,  UAEH, carretera Pachuca-Tulancingo km 4.5, C.P. 42184 , Pachuca, Hidalgo, M\'exico}

\begin{abstract}

The expression of the impact parameter, in the analysis for classical and semiclassical scattering cross sections for black holes, is obtained with nonlinear electrodynamics (NLED) while the absorption section is studied with the sinc approximation in the eikonal limit considering  NLED.  As an illustration, we calculate the classical and semiclassical scattering as well as the absorption sections for three black holes under NLED, the regular magnetic Bardeen and Bronnikov black holes and the singular Euler-Heisenberg black hole.  All are compared with the sections of their linear electromagnetic counterpart, the Reissner-Nordstr\"{o}m (RN) black hole.
The comparison shows how NLED affects the sections as well as the variation with respect to the Reissner-Nordstr\"{o}m sections, in some cases these variations are small.

\end{abstract}


\maketitle

\section{Introduction}

In general relativity (GR) the black holes (BH) are objects of much study because they are simple objects that can be described by their mass, angular momentum and charge \cite{Townsend:1997ku}. Also, its physical implications, as they behave like thermodynamic systems possessing temperature and entropy. Another important implication is the presence of singularities in the BH spacetime.

A way to study the physics of BH is to analyse test fields around them. For example an important aspect of BHs is to study the absorption and scattering  of matter and fields around them, because the dynamics of a BH can be explored as stability and gravitational wave emission. In that sense analyzing the  scattering and absorption sections is important,  this has been studied in many BH scenarios. \cite{Dolan:2008kf} \cite{Sebastian:2014cka} \cite{Macedo:2015ikq}.

For the classical approximation of the scattering is possible to consider a stream of parallel null geodesics coming from infinity with a critical impact parameter \cite{Collins:1973xf}. Also, the semiclassical glory approximation \cite{PhysRevD.31.1869} could be used because this considers the interference that occurs between scattered partial waves with different angular momenta.

Another important aspect to study is when the test field is absorbed by the BH. The absorption cross-section can be used when the low-frequency limit of a massless neutral scalar field is equal to the area of the horizon \cite{Higuchi:2001si}. In the high-frequency limit (sinc approximation) the absorption cross-section is proportional to the sum of the called geometric cross-section and oscillatory part of the absorption cross-section  \cite{Decanini:2011xi}.

In the eikonal limit, the absorption cross-section can be written in terms of the parameters of the unstable null circular orbits as the angular velocity and the Lyapunov exponent.

In the presence of nonlinear electrodynamics field (NLED), the behavior of photons and massless test particles are not the same. The photons do propagate along null geodesics of an effective geometry that depends on the nonlinear theory. Then, the electromagnetic fields modify the trajectories of the null geodesics \cite{Gutierrez:1981ed}.  The above would imply that in the aforementioned limits there would be modifications of the absorption and dispersion sections. In \cite{Breton:2016mqh} the quasinormal modes of black holes were studied with nonlinear electrodynamics in the eikonal approximation and the way the effective metric modifies the equation of motion for a test particle in the static spherically symmetric (SSS) space-time and expressions the Lyapunov exponent.

For example, in \cite{Breton:2017hwe}, it was studied QNM modes and absorption cross-sections of Born-Infeld-de Sitter black holes via the sinc approximation, considering the effects of (NLED).

The proposal of the paper focuses on studying how the effects of nonlinear electrodynamics modify the classical and semi-classical scattering section as well as the adsorption section in the sinc approximation. The black holes of Bardeen, Bronnikov, and Euler-Heisenberg are considered as examples because they are black holes in non-linear electrodynamics and they are compared with Reissner-Nordstrom (RN) black hole that is the linear counterpart of nonlinear electromagnetic black holes.

The paper is organized as follows. In Sec. II, a summary of the nonlinear electrodynamics and the effective metric is presented. In III, the expressions for the classical and semiclassical scattering sections in terms of the effective metric are presented.  In Sect. IV, we describe how the adsorption section can be obtained by sinc approximation and considering the effects of NLED. The scattering and absorption sections of some black holes in NLED are compared with the sections of Reissner-Nordstr\"{o}m in Section V. Finally, the conclusions are given in the last section.

\section{ effective metric}

The action for  gravitation  coupled to  a nonlinear electromagnetic (NLED) field is given by;

\begin{equation}\label{action}
S=\frac{1}{16\pi}\int d^{4}x\sqrt{-g}[R-L(F,G)],
\end{equation}
where $R$ is the scalar curvature and $L$ is an arbitrary function of the electromagnetic invariant  $F=F^{\mu\nu}F_{\mu\nu}$  ,  with $F_{\mu\nu}=\partial_{\mu}A_{\nu}-\partial_{\nu}A_{\mu}$ being  the electromagnetic field tensor. In this work, we consider solutions with only electric or magnetic charges, although the most general action can depend on both charges (see \cite{Hendi:2013ika}).

When considering a statical spherically symmetric metric;

\begin{equation}\label{sss}
ds^{2}=f(r)dt^{2}-\frac{1}{f(r)}dr^{2}-r^{2}(d\Theta^{2}+\sin^{2} \Theta d \phi^{2}),
\end{equation}

, the two nonzero components of the electromagnetic tensor are:
\begin{equation}
r^{2}L_{F}F^{10}=q_{e}, \quad  F_{23}=q_{m}\sin \Theta,
\end{equation}
where $q_{e}$ and $q_{m}$ are the electric and magnetic charges.

The photons propagate along null geodesics of an effective geometry which depends on the nonlinear theory \cite{Plebanski:106680}, that is, the background metric $g_{\mu\nu}$  is replaced by an effective metric $\gamma^{\mu\nu}_{\rm eff}$;

\begin{equation}
\gamma_{\mu\nu}=L_{F}g^{\mu\nu}-4L_{FF}F^{\mu}_{\alpha}F^{\alpha\nu}
\end{equation}

The calculation of the effective metric components $\gamma^{\mu\nu}_{\rm eff}$ of the SSS spacetime were obtained in \cite{Breton:2016mqh} for the case of a static spacetime;

\begin{equation}\label{efec}
ds_{\rm eff}^{2}=(L_{F}G_{m})^{-1}\left\{G_{m}G_{e}^{-1}\left( f(r)dt^{2} - \frac{1}{f(r)}dr^{2}\right)-r^{2}(d\Theta^{2}+\sin^{2} \Theta d \phi^{2})\right\}.
\end{equation}

$G_m$ and $G_e$, the magnetic and electric factors that make the difference between the linear and nonlinear electromagnetism, are given by

\begin{equation}\label{Gs}
G_{m}=\left(1+4L_{FF}\frac{q_{m}^{2}}{L_{F}r^{4}}\right) , \;\;\;\;\;\;\ G_{e}=\left(1-4L_{FF}\frac{q_{e}^{2}}{L_{F}^{3}r^{4}}\right);
\end{equation}
The subindex $F$ means the first or second derivative with respect to $F$ and in the linear limit become equal to one.

\section{Scattering section}
To investigate the scattering it's necessary to find the scattering differential cross-section, a  procedure used is the study of geodesics for the classical approximation. The geodesic analysis is important because at very high frequencies the wave propagates along of null geodesics \cite{Collins_1973}. The photons propagate along null geodesics is described by the following Lagrangian density:

\begin{equation}\label{lagrange}
\mathfrak{L}= \frac{1}{2}\gamma^{\mu\nu}_{\rm eff}S_{, \mu}S_{, \nu}=0.
\end{equation}

When we consider the massless particles the Lagrangian density is $\mathfrak{L}= \frac{1}{2}S_{, \mu}S_{, \nu}=0$. We restrict the geodesic motion to the plane $\Theta=\pi / 2$. The energy  $E=\frac{\partial \mathfrak{L}}{\partial \dot{t}}$ (the overdot denotes differentiation with respect to an affine parameter )  and the angular momentum $L=\frac{\partial \mathfrak{L}}{\partial \dot{\phi}}$ of a test particle  are conserved quantities,

\begin{equation}\label{constan}
G_{m}G_{e}^{-1}f(r)\dot{t}=  E= {\rm const},\;\;\;\;\;\;\;\ r^{2}\dot{\phi}=  L= {\rm const}.
\end{equation}

From equation (\ref{lagrange}) and the line element (\ref{efec}) we can obtain for null geodesics;

\begin{equation}\label{geo}
\left(\frac{du}{d\phi}\right)^2=\left(\frac{G_e}{G_m}\right)^2\left(\frac{1}{b^2}-u^2f(1/u)\frac{G_m}{G_e}\right).
\end{equation}

where $u=1/r$ and  $b=L/E$ is the impact parameter in terms of the constants of motion. Now differentiation with respect to $\phi$, yields;

\begin{eqnarray}\label{geo1}
\frac{d^2u}{d\phi^2}= \frac{G_e}{G_m}  \frac{d}{du}\left(\frac{1}{b^2}-u^2 f(1/u) \frac{G_m}{G_e}\right)+ \left(\frac{G_e}{G_m}\right)^2\left(-2uf(1/u)\frac{G_m}{G_e} \right.\nonumber\\ - u^2  \left.\frac{d f(1/u)}{du}\frac{G_m}{G_e}- u^2f(1/u)\frac{d}{du}\frac{G_m}{G_e}\right).
\end{eqnarray}

By solving the last equations (\ref{geo}) and (\ref{geo1}), the impact parameter from the critical orbits is obtained,  i.e we consider the smallest positive root $u_{c}$ of (\ref{geo}) and substituting its value in (\ref{geo1}) the impact parameter $b_{c}$ is obtained. For any $b>b_c$ there is no scattering, so the frontier conditions used while solving (\ref{geo}) and (\ref{geo1}) must consider this information.

The deflection angle is given by;

\begin{equation}\label{impact parameter}
\theta(b)=2\alpha(b)-\pi
\end{equation}
where
\begin{equation}
 \alpha=\int_0^{u_0}du \frac{G_m}{G_e}\left(\frac{1}{b^2}-u^2 f(1/u) \frac{G_m}{G_e}\right)^{-1/2},
\end{equation}

 which is obtained from the integration of (\ref{geo}) where $u_{0}$ is the turning point ($u_{0}=1/r_{0}$).

Finally, with the impact parameter $b(\theta)$, associated with a scattering angle $\theta$, the scattering section is given by;

 \begin{equation}
 \frac{d\sigma}{d\Omega}=\frac{1}{\sin \theta} \sum b(\theta) |\frac{db(\theta)}{d \theta}|
 \end{equation}.

 Another procedure to investigate the scattering differential cross-section, is the semiclassical approach (Glory scattering) \cite{PhysRevD.31.1869} where the geodesics can be seen as waves. This captures the waves interference for large scattering angles. For Glory scattering, the differential cross-section by spherically symmetric Black Holes is given  by;

\begin{equation}
\frac{d\sigma_g}{d\Omega}=2\pi\omega b_g^2 \left| \frac{db}{d\theta}\right|_{\theta=\pi}J_{2s}^2(\omega b_g \sin\theta)
\end{equation}

with $\omega$ as the wave frequency. $J_{2s}^2(x)$ stands for the Bessel function of first kind of order $2s$ where $s$ represents the spin, $s=0$ for scalar waves. The impact parameter of the reflected waves ($\theta \sim\pi$) is denoted by $b_g$ where $b$ is the impact parameter (\ref{impact parameter}). As a semiclassical approximation, it is valid for $M\omega \gg 1$ ($M$ the mass of the BH).

\section{Absorption section}

In the case of  mid-to-high frequencies, the sinc approximation can be used  to analyze the absorption cross-section as in \cite{PhysRevD.83.044032},  where the high-energy absorption cross-section is studied in the eikonal regime, from the characteristics of the null unstable geodesics.

Then  the absorption cross-section  in the eikonal limit is defined by;

\begin{equation}
\sigma_{abs}=\pi b_{c}^{2} - 4\pi \frac{\lambda b_{c}^{2}}{\omega}e^{-\pi \lambda b_{c}}\sin(2\pi \omega b_{c})
\end{equation}

 where $\lambda$ is the Lyapunov exponent and $b_c$ is the critical impact parameter.

As discussed in \cite{Breton:2016mqh} the exponent is modified  for the effective metric (\ref{efec}) in the (NLED) as:

\begin{equation}\label{expLyapunov1}
\lambda^{2}=\frac{f_{c}r^{2}_{c}}{2}\left[  \frac{f_{c}}{r_{c}^{2}} \frac{G_{m}}{G_{e}} \left( \frac{G_{e}}{G_{m}}\right)^{''}_{c}- \left(\frac{f}{r^{2}}\right)^{''}_{c}\right],
\end{equation}

In circular unstable orbits, the impact parameter, when subjected to NLED is given by;
\begin{equation}
b_{c}=\sqrt{\frac{G_e}{G_m}\frac{r_{c}^{2}}{f(r)_{c}}},
\end{equation}

in this limit the quasinormal modes are  approximated as;

\begin{equation}
\omega_{n} \sim b^{-1}_{c}- i (n+1/2)|\lambda|.
\end{equation}

The idea of considering the effective metric in nonlinear electrodynamics by obtaining the absorption cross sections in the high frequency limit via the sinc approximation was addressed in \cite{Breton:2017hwe} in the study of Born-Infeld-de Sitter black holes.

\section{Examples}

We calculate the scattering differential and absorption sections for black holes in NLED theory, considering the modification of the effective metric. Then, we analyze and compare the differences with the case when the effective metric is not considered (for massless particles). Also, the different sections of the black holes can be similar but not identical to the sections of Reissner-Nordstr\"{o}m.

\subsection{Bardeen black hole}

 The Bardeen BH can be interpreted as a self-gravitating  nonlinear magnetic monopole \cite{AyonBeato:2000zs} with mass $M$ and magnetic charge $q_{m}=g$,  derived from  Einstein gravity coupled to the nonlinear Lagrangian,

\begin{equation}
L(F)=\frac{6}{s g^{2}}\frac{(g^{2}F/2)^{\frac{5}{4}}}{(1+\sqrt{g^{2}F/2})^{\frac{5}{2}}},
\label{ABGlag}
\end{equation}

where  $s= |g|/2M$.  The solution for the coupled Einstein  and NLED Lagrangian (\ref{ABGlag}) for a static spherically symmetric space  is given by,

\begin{equation}\label{Bardeenmetric1}
f(r)=1-\frac{2Mr^{2}}{(r^{2}+g^{2})^{\frac{3}{2}}}.
\end{equation}

The solution (\ref{Bardeenmetric1}) has horizons only if $2s= g/m \leq 0.7698$. The electromagnetic invariant is $F=2g^{2}/r^{4}$ and the $G_e$ and $G_m$ factors are  given by

\begin{equation}
G_{e}=1, \quad G_{m}=1-\frac{4(6g^{2}-r^{2})}{8(r^{2}+g^{2})}.
\label{GsBardeen}
\end{equation}

When we study instability, we find that the horizon radius, $r_{c}>r_{h}$. In \cite{Macedo:2017szm} \cite{Huang:2014nka}, absorption cross sections were studied for a massless scalar wave in Bardeen black hole, also in \cite{Macedo:2015qma}, the cross-section was computed.

 We shall focus on the absorption and scattering cross-section of the Bardeen black hole. To carry out the study, we have used $r\rightarrow rM$  and $g\rightarrow \frac{4M}{3\sqrt{3}}g$.

\begin{figure}[h]
\begin{center}
\includegraphics [width =0.5 \textwidth ]{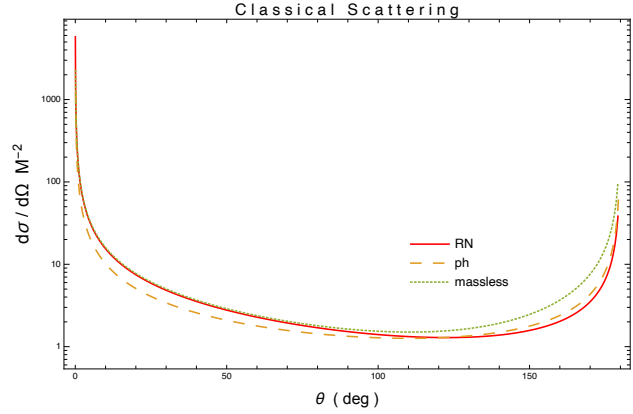}
\end{center}
\caption{Behavior of classical scattering cross-section of  Bardeen for fixed values of parameters $g= 0.5$ and $M\omega =2$ }\label{Fig1}
\end{figure}

In figure \ref{Fig1} the classical scattering sections for Bardeen and RN black holes are compared  (see \cite{Crispino:2009ki}\cite{PhysRevD.89.104053}). It can be observed that for massless particles, the classical section is higher than the one obtained for photons in Bardeen black hole. It can also be noted that for small angles there is not a significant difference between RN and Bardeen massless scattering sections. It is worth mentioning that the massless section is higher than the one obtained for photons in Bardeen black hole and for RN.

\begin{figure}[h]
\begin{center}
\includegraphics [width =0.5 \textwidth ]{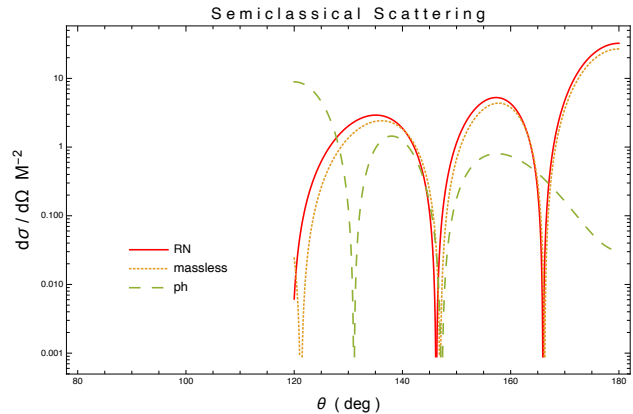}
\end{center}
\caption{Behavior of semiclassical scattering cross-section of  Bardeen  for fixed values of parameters $g= 0.5$ and  $M\omega =2$ }\label{Fig2}
\end{figure}

Figure \ref{Fig2} shows the semi classical scattering sections for Bardeen and RN black holes. It can be observed that in the case with massless particles the scattering section is highly similar to RN section, however, by considering photons in Bardeen black hole significant differences arise. The limit $g \rightarrow 0$  $G_{m} \rightarrow 3/2$ could be the reason that such differences exist.

\begin{figure}[h]
\begin{center}
\includegraphics [width =0.5 \textwidth ]{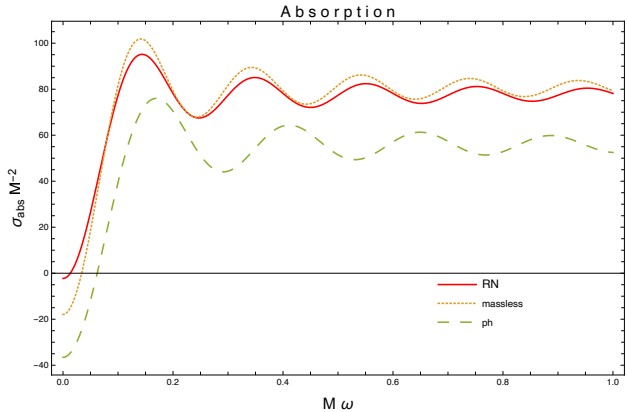}
\end{center}
\caption{Behavior of absorption section of  Bardeen  for fixed values of parameters $g= 0.5$}\label{Fig3}
\end{figure}

The figure \ref{Fig3} shows the absorption sections for Bardeen and RN black holes. It can be observed that when considering massless particles in Bardeen black holes the absorption section is highly similar to RN, such as was reported in \cite{PhysRevD.90.064001}. When considering photons, the Bardeen section is lower than the massless and RN sections.

\subsection{Bronnikov  black hole}

The Lagrangian for the magnetic black holes of Bronnikov  \cite{Bronnikov:2000vy}  is;
\begin{equation}
L(F)=F {\rm sech}^2 [a(F/2)^{1/4}],
\end{equation}

where $a$ is a constant. The metric function in the line element of the form (\ref{sss}) is

\begin{equation}\label{Bronnokovmetric}
f(r)=1-\frac{g^{3/2}}{ar} \left[ {1- \tanh \left( \frac{a \sqrt{g}}{r}\right)} \right],
\end{equation}
where the constant $a$ is related to the mass $m$ and the magnetic charge $g$ by $a=g^{3/2}/(2M)$. The solution corresponds to a BH if $M/g > 0.96$. The electromagnetic invariant  is $F=2g^{2}/r^{4}$. For a purely magnetic charge $G_{e}=1$ and;

\begin{equation}
 G_{m}=\frac{g^{2}\sinh^{2}(\frac{g^{2}}{2Mr})\left(-2g^{2}+g^{2}\cosh(\frac{g^{2}}{2Mr})-5Mr\sinh(\frac{g^{2}}{2Mr})\right)}{4Mr\left(-4Mr^{3}+g^{2}\tanh(\frac{g^{2}}{2Mr})\right)}
\end{equation}

To study the behavior of the scattering and absorption section of the black hole of Bronnikov, we consider $r\rightarrow rM$  and $g\rightarrow \frac{M}{0.96}g$, the sections are compared with RN.

\begin{figure}[h]
\begin{center}
\includegraphics [width =0.5 \textwidth ]{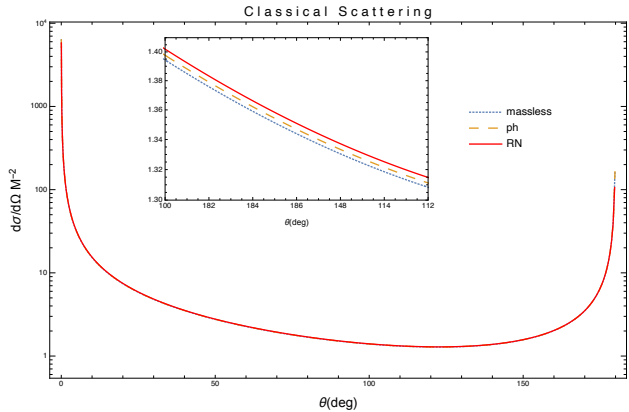}
\end{center}
\caption{Behavior of classical scattering cross-section of  Bronnikov  for fixed values of parameters $g= 0.5$ and $M\omega =2$ }\label{Fig4}
\end{figure}

Figure \ref{Fig4} shows the classical scattering section for Bronnikov and RN black holes. The differences while considering photons and massless particles are not considerable, in the same way, the same conclusion arises when comparing with RN, however, it is possible to compare this differences by normalizing the Bardeen sections with respect to RN, obtaining  $ (\frac{d\sigma}{d\Omega})_{RN}> (\frac{d\sigma}{d\Omega})_{photons}>(\frac{d\sigma}{d\Omega})_{massless}$.

\begin{figure}[h]
\begin{center}
\includegraphics [width =0.5 \textwidth ]{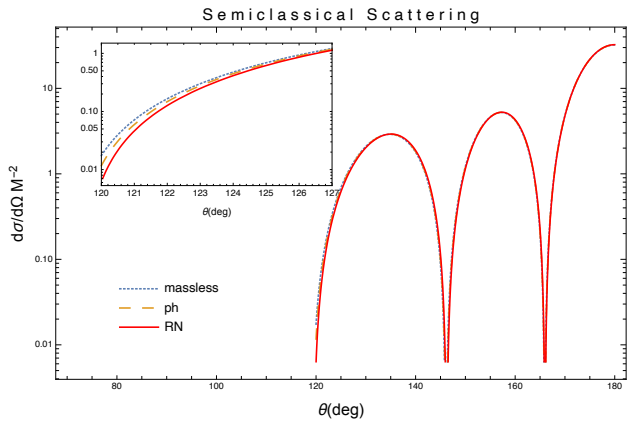}
\end{center}
\caption{Behavior of semiclassical scattering cross-section of  Bronnikov  for fixed values of parameters $g= 0.5$ and  $M\omega =2$ }\label{Fig5}
\end{figure}

In the case of the semiclassical approach, we observe that while both scattering sections (considering photons and particles massless ) are similar and very close, compared to the scattering section of RN, there are differences, but their values are not very far from the values of RN. This is shown in figure \ref{Fig5}.

\begin{figure}[h]
\begin{center}
\includegraphics [width =0.5 \textwidth ]{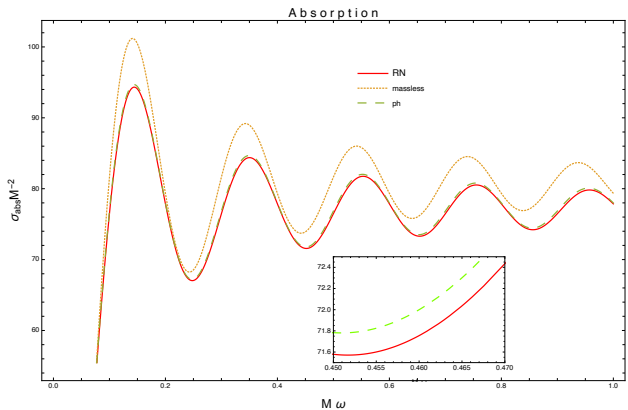}
\end{center}
\caption{Behavior of absorption section of  Bronnikov  for fixed values of parameters $g= 0.5$.}\label{Fig6}
\end{figure}

For the Bronnikov absorption section (see figure \ref{Fig6}) it is possible to observe that the case of considering photons is highly similar to the absorption section of RN while the values of the absorption section considering particles massless are higher.

Comparing the scattering sections from Bardeen black hole and  Bronnikov black hole, we observe that  the classical scattering for massless particles is greater than the one from photons for Bardeen black hole while in the case of  Bronnikov an inverse behavior is found. In the case of absorption section, the one found for massless particles is greater than one for photons in the Bronnikov black hole while for Bardeen, the opposite case is found.

It would be adequate to observe the behavior of a non-regular black hole to compare the resulting sections with the ones obtained from regular black holes.

\subsection{Euler-Heisenberg Black hole}

The lagrangian for the magnetic  black hole the Euler and  Heisenberg was obtained of  the effective action for electrodynamics due to one-loop quantum corrections \cite{Yajima:2000kw}, and is given by;

\begin{equation}\label{EHlagr}
L(F)= F(1-aF),
\end{equation}

with $a=he^{2}/(360\pi^{2}m_{e}^{2})$, where $h$, $e$, and $m_{e}$ are the Planck constant, electron charge, and electron mass respectively.

The corresponds  metric  elements in  (\ref{sss}) given by;

\begin{equation}
f(r)=g(r)=1-\frac{2M}{r}+\frac{q^{2}}{r^{2}}-\frac{2}{5}a\frac{q^{4}}{r^{6}},
\end{equation}

where $M$ is the mass parameter and $q$ is the magnetic charge.  One or two horizons may occur: if  $(M/q)^{2}\leq 24/25$ a single horizon exists but for  $(M/q)^{2}> 24/25$ a second and a third horizons occur. The electromagnetic invariant is  $F={2q^{2}}/{r^{4}}$.  The $G_e$ and $G_m$ factors are;

\begin{equation}
G_{e}=1, \quad G_{m}=\left(1+\frac{8aq^{2}}{4aq^{2}-r_c^{4}}\right).
\end{equation}

As in the cases shown before, we consider the following $r\rightarrow M r$, $a\rightarrow \frac{8}{27}q^2$ and $q\rightarrow \frac{5 M}{\sqrt{24}}q$, then it is possible to obtain the classical scattering section.

\begin{figure}[h]
\begin{center}
\includegraphics [width =0.5 \textwidth ]{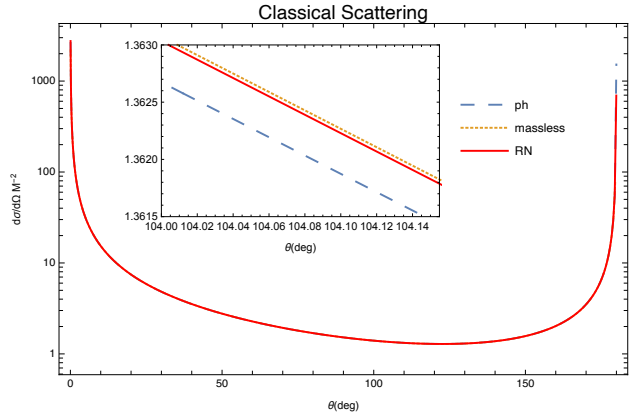}
\end{center}
\caption{Behavior of classical scattering cross-section of Euler-Heisenberg  for fixed values of parameters $q= 0.5$, $a=0.8$ and $M\omega =2$ }\label{Fig7}
\end{figure}

It can be noted that when massless particles are considered, the behavior remains near RN black hole as can be noted in figure \ref{Fig7}, while the deviation with respect to the photon case is more significant.

\begin{figure}[h]
\begin{center}
\includegraphics [width =0.5 \textwidth ]{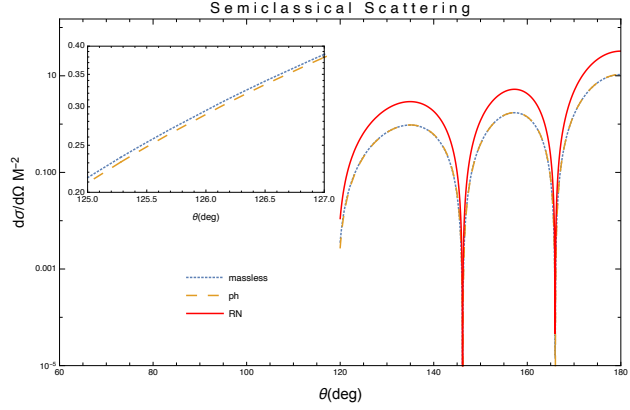}
\end{center}
\caption{Behavior of semiclassical scattering cross-section of  Euler-Heisenberg  for fixed values of parameters $q= 0.5$, $a=0.8$ and  $M\omega =2$ }\label{Fig8}
\end{figure}

Now, in figure \ref{Fig8}, the behavior for the semiclassical scattering section is shown. In this case, the semiclassical section for photons does not show significant differences with the case of massless particles, however, both show  differences with respect to the RN black hole.

\begin{figure}[h]
\begin{center}
\includegraphics [width =0.5 \textwidth ]{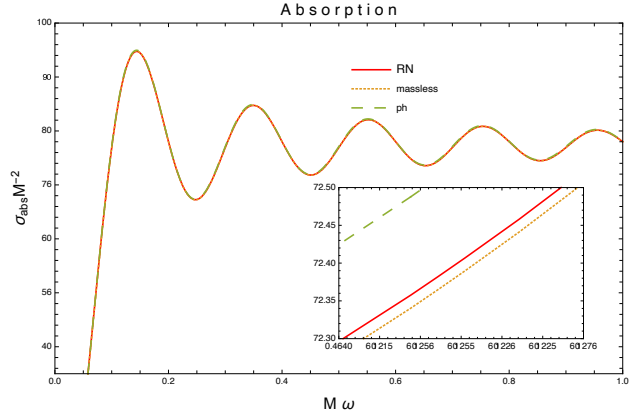}
\end{center}
\caption{Behavior of absorption section of  Euler-Heisenberg  for fixed values of parameters $q= 0.5$, $a=0.8$. }\label{Fig9}
\end{figure}

Finally, we studied the absorption section (see figure \ref{Fig9}), when we consider the photons, the behavior shows that differences are present in lesser scale than the regular Bardeen black hole. However, in the Euler Heisenberg case, the absorption section is more similar to RN than the Bardeen case.

\section{Conclusions}

In this paper, it is first shown how the scattering and absorption sections are modified by considering the effective metric generated by the action of NLED for null geodesics. In particular, by considering the classical and semiclassical scattering, the effect from NLED is reflected in the impact parameter $b$, which is modified in function of the electric and magnetic factors from the effective metric ($G_e$ and $G_m$ respectively) as well as their derivatives.

In the absorption case, the sinc approximation is used, so that the absorption section is given in function of the Lyapunov exponent and the impact parameter for a null circular unstable geodesic, which as discussed in \cite{Breton:2016mqh}, it was also modified by considering the effective metric.

To show the effects of considering the modification of the impact parameter generated by NLED, the scattering and absorption sections from two regular black holes and a non-regular one were studied and analyzed.

The Bardeen and Bronnikov regular black holes were considered to study the effects that the effective metric generates over the scattering and absorption sections. It can be seen that in the case of classical and semiclassical scattering the one corresponding to Bronnikov is closer to the RN section. However, in the Bronnikov case, the difference between the scattering sections from massless particles and photons are minimal. When studying the absorption section, it is found that in the Bronnikov case of photons the difference with RN is minimal while for Bardeen the difference is significative. This behavior is opposite when considering massless particles.

In the case of the Euler Heisenberg non regular black hole, the classical scattering section  for photons is closer to the RN. By using the semiclassical approximation, the effects due to NLED are notable. Finally, when studying the sinc approximation for EH, it is possible to note that, unlike the regular black holes, in this case, the absorption section is greater for photons.

\bibliographystyle{unsrt}

\bibliography{bibliografia}

\end{document}